\documentclass[12pt]{article}

\usepackage{graphicx}
\usepackage{enumitem}
\usepackage{float}
\usepackage{amssymb}
\usepackage{comment}
\usepackage{mathtools}
\input{epsf}

\advance \topmargin by -\headheight
\advance \topmargin by -\headsep

\evensidemargin \oddsidemargin
\begin{document}

\topmargin -.6in

\def\rh{{\hat \rho}}
\def\alie{{\hat{\cal G}}}
\newcommand{\sect}[1]{\setcounter{equation}{0}\section{#1}}
\renewcommand{\theequation}{\thesection.\arabic{equation}}

\def\rf#1{(\ref{eq:#1})}
\def\lab#1{\label{eq:#1}}
\def\nonu{\nonumber}
\def\br{\begin{eqnarray}}
\def\er{\end{eqnarray}}
\def\be{\begin{equation}}
\def\ee{\end{equation}}
\def\eq{\!\!\!\! &=& \!\!\!\! }
\def\foot#1{\footnotemark\footnotetext{#1}}
\def\lb{\lbrack}
\def\rb{\rbrack}
\def\llangle{\left\langle}
\def\rrangle{\right\rangle}
\def\blangle{\Bigl\langle}
\def\brangle{\Bigr\rangle}
\def\llbrack{\left\lbrack}
\def\rrbrack{\right\rbrack}
\def\lcurl{\left\{}
\def\rcurl{\right\}}
\def\({\left(}
\def\){\right)}
\newcommand{\nit}{\noindent}
\newcommand{\ct}[1]{\cite{#1}}
\newcommand{\bi}[1]{\bibitem{#1}}
\def\lskip{\vskip\baselineskip\vskip-\parskip\noindent}
\relax

\def\tr{{\rm Tr}}
\def\Tr{\mathop{\rm Tr}}
\def\trace{\widehat{\rm Tr}}
\def\v{\vert}
\def\bv{\bigm\vert}
\def\Bgv{\;\Bigg\vert}
\def\bgv{\bigg\vert}
\newcommand\partder[2]{{{\partial {#1}}\over{\partial {#2}}}}
\newcommand\funcder[2]{{{\delta {#1}}\over{\delta {#2}}}}
\newcommand\Bil[2]{\Bigl\langle {#1} \Bigg\vert {#2} \Bigr\rangle}  
\newcommand\bil[2]{\left\langle {#1} \bigg\vert {#2} \right\rangle} 
\newcommand\me[2]{\left\langle {#1}\bv {#2} \right\rangle} 
\newcommand\sbr[2]{\left\lbrack\,{#1}\, ,\,{#2}\,\right\rbrack}
\newcommand\pbr[2]{\{\,{#1}\, ,\,{#2}\,\}}
\newcommand\pbbr[2]{\lcurl\,{#1}\, ,\,{#2}\,\rcurl}

\def\ket#1{\mid {#1} \rangle}
\def\bra#1{\langle {#1} \mid}
\newcommand{\braket}[2]{\langle {#1} \mid {#2}\rangle}
%
\def\a{\alpha}
\def\at{{\tilde A}^R}
\def\atc{{\tilde {\cal A}}^R}
\def\atcm#1{{\tilde {\cal A}}^{(R,#1)}}
\def\b{\beta}
\def\dc{{\cal D}}
\def\d{\delta}
\def\D{\Delta}
\def\eps{\epsilon}
\def\vareps{\varepsilon}
\def\g{\gamma}
\def\G{\Gamma}
\def\grad{\nabla}
\def\h{{1\over 2}}
\def\l{\lambda}
\def\L{\Lambda}
\def\m{\mu}
\def\n{\nu}
\def\o{\over}
\def\om{\omega}
\def\O{\Omega}
\def\p{\phi}
\def\P{\Phi}
\def\pa{\partial}
\def\pr{\prime}
\def\pt{{\tilde \Phi}}
\def\qs{Q_{\bf s}}
\def\ra{\rightarrow}
\def\s{\sigma}
\def\S{\Sigma}
\def\t{\tau}
\def\th{\theta}
\def\Th{\Theta}
\def\tpp{\Theta_{+}}
\def\tmm{\Theta_{-}}
\def\tpg{\Theta_{+}^{>}}
\def\tms{\Theta_{-}^{<}}
\def\tp0{\Theta_{+}^{(0)}}
\def\tm0{\Theta_{-}^{(0)}}
\def\ti{\tilde}
\def\wti{\widetilde}
\def\jc{J^C}
\def\bj{{\bar J}}
\def\sj{{\jmath}}
\def\bsj{{\bar \jmath}}
\def\bp{{\bar \p}}
\def\vp{\varphi}
\def\ve{\varepsilon}
\def\vt{{\tilde \varphi}}
\def\faa{Fa\'a di Bruno~}
\def\ca{{\cal A}}
\def\cb{{\cal B}}
\def\ce{{\cal E}}
\def\cg{{\cal G}}
\def\cgh{{\hat {\cal G}}}
\def\ch{{\cal H}}
\def\chh{{\hat {\cal H}}}
\def\cl{{\cal L}}
\def\cm{{\cal M}}
\def\cn{{\cal N}}
\def\u2{\mid u\mid^2}
\def\ub{{\bar u}}
\def\z2{\mid z\mid^2}
\def\zb{{\bar z}}
\def\w2{\mid w\mid^2}
\def\wb{{\bar w}}
\newcommand\sumi[1]{\sum_{#1}^{\infty}}   
\newcommand\fourmat[4]{\left(\begin{array}{cc}  
{#1} & {#2} \\ {#3} & {#4} \end{array} \right)}
\newcommand{\re}[1]{(\ref{#1})}

%
\def\lie{{\cal G}}
\def\kmlie{{\hat{\cal G}}}
\def\dlie{{\cal G}^{\ast}}
\def\elie{{\widetilde \lie}}
\def\edlie{{\elie}^{\ast}}
\def\hlie{{\cal H}}
\def\flie{{\cal F}}
\def\wlie{{\widetilde \lie}}
\def\f#1#2#3 {f^{#1#2}_{#3}}
\def\winf{{\sf w_\infty}}
\def\win1{{\sf w_{1+\infty}}}
\def\hwinf{{\sf {\hat w}_{\infty}}}
\def\Winf{{\sf W_\infty}}
\def\Win1{{\sf W_{1+\infty}}}
\def\hWinf{{\sf {\hat W}_{\infty}}}
\def\Rm#1#2{r(\vec{#1},\vec{#2})}          
\def\OR#1{{\cal O}(R_{#1})}           
\def\ORti{{\cal O}({\widetilde R})}           
\def\AdR#1{Ad_{R_{#1}}}              
\def\dAdR#1{Ad_{R_{#1}^{\ast}}}      
\def\adR#1{ad_{R_{#1}^{\ast}}}       
\def\KP{${\rm \, KP\,}$}                 
\def\KPl{${\rm \,KP}_{\ell}\,$}         
\def\KPo{${\rm \,KP}_{\ell = 0}\,$}         
\def\mKPa{${\rm \,KP}_{\ell = 1}\,$}    
\def\mKPb{${\rm \,KP}_{\ell = 2}\,$}    
%
\def\rlx{\relax\leavevmode}
\def\inbar{\vrule height1.5ex width.4pt depth0pt}
\def\IZ{\rlx\hbox{\sf Z\kern-.4em Z}}
\def\IR{\rlx\hbox{\rm I\kern-.18em R}}
\def\IC{\rlx\hbox{\,$\inbar\kern-.3em{\rm C}$}}
\def\IN{\rlx\hbox{\rm I\kern-.18em N}}
\def\IO{\rlx\hbox{\,$\inbar\kern-.3em{\rm O}$}}
\def\IP{\rlx\hbox{\rm I\kern-.18em P}}
\def\IQ{\rlx\hbox{\,$\inbar\kern-.3em{\rm Q}$}}
\def\IF{\rlx\hbox{\rm I\kern-.18em F}}
\def\IG{\rlx\hbox{\,$\inbar\kern-.3em{\rm G}$}}
\def\IH{\rlx\hbox{\rm I\kern-.18em H}}
\def\II{\rlx\hbox{\rm I\kern-.18em I}}
\def\IK{\rlx\hbox{\rm I\kern-.18em K}}
\def\IL{\rlx\hbox{\rm I\kern-.18em L}}
\def\one{\hbox{{1}\kern-.25em\hbox{l}}}
\def\0#1{\relax\ifmmode\mathaccent"7017{#1}%
B        \else\accent23#1\relax\fi}
\def\omz{\0 \omega}
%
\def\ltimes{\mathrel{\vrule height1ex}\joinrel\mathrel\times}
\def\rtimes{\mathrel\times\joinrel\mathrel{\vrule height1ex}}
%
\def\mark{\noindent{\bf Remark.}\quad}
\def\prop{\noindent{\bf Proposition.}\quad}
\def\theor{\noindent{\bf Theorem.}\quad}
\def\name{\noindent{\bf Definition.}\quad}
\def\exam{\noindent{\bf Example.}\quad}
\def\proof{\noindent{\bf Proof.}\quad}
%

\begin{titlepage}
\vspace*{-1cm}

\vskip 3cm

\vspace{.2in}
\begin{center}
{\large\bf  False vacuum Skyrmions revisited}
\end{center}

\vspace{.5cm}

\begin{center}
 L. R. Livramento$^{\dagger }$\footnote{livramento@theor.jinr.ru}
and Ya. Shnir$^{\dagger \star }$\footnote{shnir@theor.jinr.ru}

\vspace{.3 in}
\small

\par \vskip .2in \noindent
$^{\dagger}$BLTP, JINR, Dubna 141980, Moscow Region, Russia\\

\par \vskip .2in \noindent
$^{\star}$Institute of Physics, University of Oldenburg,
Oldenburg D-26111, Germany

\normalsize
\end{center}

\vspace{.5in}

\begin{abstract}

We consider the classical  static soliton solutions of the Skyrme model with false vacuum potential. We make use of fully three-dimensional relaxation calculations
to construct global energy minimizers in the sectors of topological degrees from $Q=1$ to $Q=6$. These solutions may be metastable, they contain a domain of true vacuum inside the core. Further, we  explore small regions of negative topological charge density which
appear for the Skyrmions of degrees $Q=3,5,6$.

\end{abstract}
\end{titlepage}
\section*{Introduction}
Many nonlinear classical field theories admit solitons, they represent regular spatially localized field configuration with finite energy, see e.g.  \cite{mantonbook,Shnir:book1}. One of the celebrated examples are Skyrmions, the topological soliton solutions of the generalized non-linear sigma model in (3+1) dimensions \cite{skyrme1,skyrme2}, for a review see \cite{Zahed:1986qz,brown2010multifaceted,MantonsBook}. Originally, the Skyrmions were interpreted as nucleons, with
identification of the baryon number and the topological charge of the field configuration. This idea acquired popularity in 1980s since Witten pointed out that,  in the limit of infinite number of colours, the Skyrme model can be considered as a low energy QCD effective theory \cite{witten1,witten2}.
 Apart from being a simple example of a relativistic field model, which support topological solitons, the Skyrme theory attracted a lot of attention due to its relation to the holonomy of the Yang-Mills instantons via the Atiyah-Manton construction \cite{Atiyah:1989dq} and to the Sakai-Sugimoto model of holographic QCD \cite{Sakai:2004cn}. On the other hand, Skyrmion-type solutions naturally arise in various condensed matter planar systems with intrinsic and induced chirality \cite{Bogdanov:1989,Bogdanov:1995,Leonov2015,Nagaosa2013}.

According to the scaling arguments of the Derrick's theorem \cite{Derrick:1964ww},
a minimal version of the Skyrme model in 3+1 dimensional spacetime, which may supports stable topological solitons should include
both the quadratic in derivatives term  $L_2$ and  a term of fourth order in the derivatives $L_4$ (Skyrme term). However, this form is not so good for a candidate model of nuclear physics, in order to
make it more phenomenological suitable for description of baryons and pions one has to supplement it with a
potential \cite{Adkins:1983hy,Battye:2004rw,Kopeliovich:2005vg,Gudnason:2016cdo}.

There is a variety of soliton solutions of the Skyrme model constructed numerically over last three decades, starting from pioneering works \cite{Braaten:1989rg,Battye:1997qq,Houghton:1997kg}.
The simplest Skyrmion of topological degree $Q=1$ is spherically symmetric, Skyrmions of higher topological degrees may possess much more complicated symmetries, for example the Skyrmion of degree $Q=3$ is tetrahedrally symmetric, there are configurations with the symmetries of the dihedral group $D_n$, the extended dihedral groups $D_{nh}$ and $D_{nd}$, and the icosahedral group $I_n$.
The rational map parametrization, suggested in  \cite{Houghton:1997kg}, provides a nice geometric construction which is very successful at capturing most of the features of the Skyrmions.

Interestingly, it was observed that there are tiny domains of negative topological density for tetrahedral $Q=3$ Skyrmions, but none for $Q=2,4$ Skyrmions with  higher amount of symmetry \cite{Houghton:2001fe,Foster:2013bw,Leese:1993mc}.

Properties of the multisoliton solutions of the Skyrme model strongly depend on the choice of the potential, it yields the asymptotic decay of the field which defines the character of interaction between the solitons. Even for the model with the
usual pion mass term, the increase of the corresponding
mass parameter may strongly affect the structure of the multi-Skyrmion configurations \cite{Battye:2004rw,Battye:2006tb}.

The  binding energy of the Skyrmions is relatively high, in order to set it into correspondence with the experimentally known  binding energies
of physical nuclei, a number of modifications of the Skyrme model have been proposed \cite{Jackson:1985yz,Meissner:1986tc,Marleau:1990nh,Adam:2010fg,Sutcliffe:2011ig,Adam:2015ele,Gudnason:2016tiz,Gudnason:2017opo}.
Notably, solitons of the original Skyrme model do not attain the topological bound, which yields a linear relation between
the static energy of the solitons and their topological
charges $Q$. To approach this bound one has to modify the model, preserving its
topological properties \cite{Adam:2010fg,Sutcliffe:2011ig,Adam:2015ele,sut_naya_1}. An example is a truncated Skyrme model with only a sixth-order derivative term, which
is the topological current density squared, and a potential   \cite{Adam:2010fg,Adam:2015ele}, or modifications of the original theory to the form which supports self-dual equations \cite{shnir,Ferreira2017,luiz1}.
Multisoliton solutions of this reduced
model exactly saturate the topological bound; they may
interact only elastically and the self-dual multi-Skyrmion configuration resembles the system of liquid drops.

In order to construct weakly bounded multi-Skyrmion configurations one can consider a combination of the
repulsive and attractive potentials \cite{Gillard:2015eia,Gudnason:2016tiz,Gudnason:2018jia}, the latter can be represented by
the pion mass term, or  by the double vacuum potential \cite{Adam:2012sa,Perapechka:2017yyc}.
In such a case,
the repulsive part of the potential separates the constituents of the  configuration which resembles a
loosely bound collection of almost isolated spherically symmetric unit charge Skyrmions.
Furthermore, various symmetry-breaking
potentials were considered to construct half-Skyrmions \cite{Jaykka:2010bq,Gudnason:2015nxa}.

An interesting possibility is to consider a potential of the Skyrme model, which possesses both true and false vacua   \cite{Dupuis:2018utr,Ferreira:2021ryf}.
Presence of the false vacuum may affect the properties of the solitons, they become meta-stable \cite{Dupuis:2018utr,Kumar:2010mv,Lee:2013ega}, the collisions of the solitons could induce the decay of the false vacuum \cite{Gomes:2018heu} and various radiative effects \cite{Dorey:2021mdh}.
However, the previous analyse of the multisoliton solutions of the false vacuum Skyrme model was restricted to an effective theory related to the rational map approximation.
Therefore, re-examination of the results obtained in the paper \cite{Dupuis:2018utr}
seems to be warranted.

In this paper, we will study classically stable multisoliton solutions of the Skyrme model with the false vacuum potential, discussed earlier in \cite{Dupuis:2018utr}. In particular, we investigate Skyrmions of higher degrees and examine the regions of the negative topological charge density, which appear for the solitons of degrees $Q=3,5,6$ and may have an important effect on false vacuum instability.

\section{False Vacuum Skyrmions}
\label{sec:model}
The Skyrme model is a Poincar{\'e} invariant,
nonlinear $SU(2)$ sigma model field theory. The basic version of the Skyrme Lagrangian includes two terms,  $L_2+L_4$, or explicitly
\be
{\cal L}_{\rm Skyrme} = -\frac{1}{2} \,\tr \(R_\mu\,R^\mu\) + \frac{1}{16}\, \tr\(\left[R_\mu,\,R_\nu\right]\,\left[R^\mu,\,R^\nu\right]\) \,,
\label{skyrme}
\ee
where we used the rescaled energy and lengh  units  $f_{\pi}/(4\,e)$ and $2/(e\,f_\pi)$, respectively\footnote{Here $f_\pi$ is the pion decay constant and $e$ is the Skyrme coupling, the parameters can be tuned to reproduce the masses of the neutron, the pions and the delta resonance, see e.g. \cite{Adkins:1983hy}.}.
The $SU(2)$ Lie algebra valued right current is
$R_\mu =\pa_\mu U\,U^{\dagger} = R_\mu^a\,\tau_a$, with $\tau_a$ being the Pauli matrices and $U$ is the so-called Skyrme field, which belongs to the $SU(2)$ Lie group. Once we impose that $U({\bf x},t)$ takes the same matrix-value at spatial infinity,
thus the Skyrme field becomes a map $U:S^3 \mapsto S^3$ from  the compactified coordinate space ${\mathbb R}^3 \cup \{\infty\}\mapsto S^3$ onto the target space $S^3$. The topological charge $Q$ corresponds to the degree of this map and can be written in the integral representation as
\be
Q=-\frac{1}{24\,\pi^2}\,\int d^3x \,\varepsilon_{ijk}\,\tr \(R_i\,R_j\,R_k\) \, . \label{chargemain}
\ee
The Lagrangian \re{skyrme} can be supplemented by symmetry breaking potential terms. The simplest choice is the usual pion mass potential
\be
V_{{\rm mass}}= m^2 \,{\rm Tr}~(\one - U) \, ,
\label{pion-pot}
\ee
it affects the qualitative shape of Skyrmions of higher degrees \cite{Battye:2004rw,Battye:2006tb}.
The dimensionless parameter $m$ is proportional to the mass of linearized excitations of the scalar field associated with the pions.

Hereafter, we are only concerned with static solutions of the Skyrme model, so we consider the energy density
\be
{\cal E}=-\frac{1}{2} \,\tr \(R_i\,R_i\) - \frac{1}{16}\, \tr\(\left[R_i,\,R_j\right]\,\left[R_i,\,R_j\right]\) +V\(U\) \, .
\label{eng_Sk}
\ee

\begin{figure}[H]
\centering
\includegraphics[scale=0.5,angle=0]{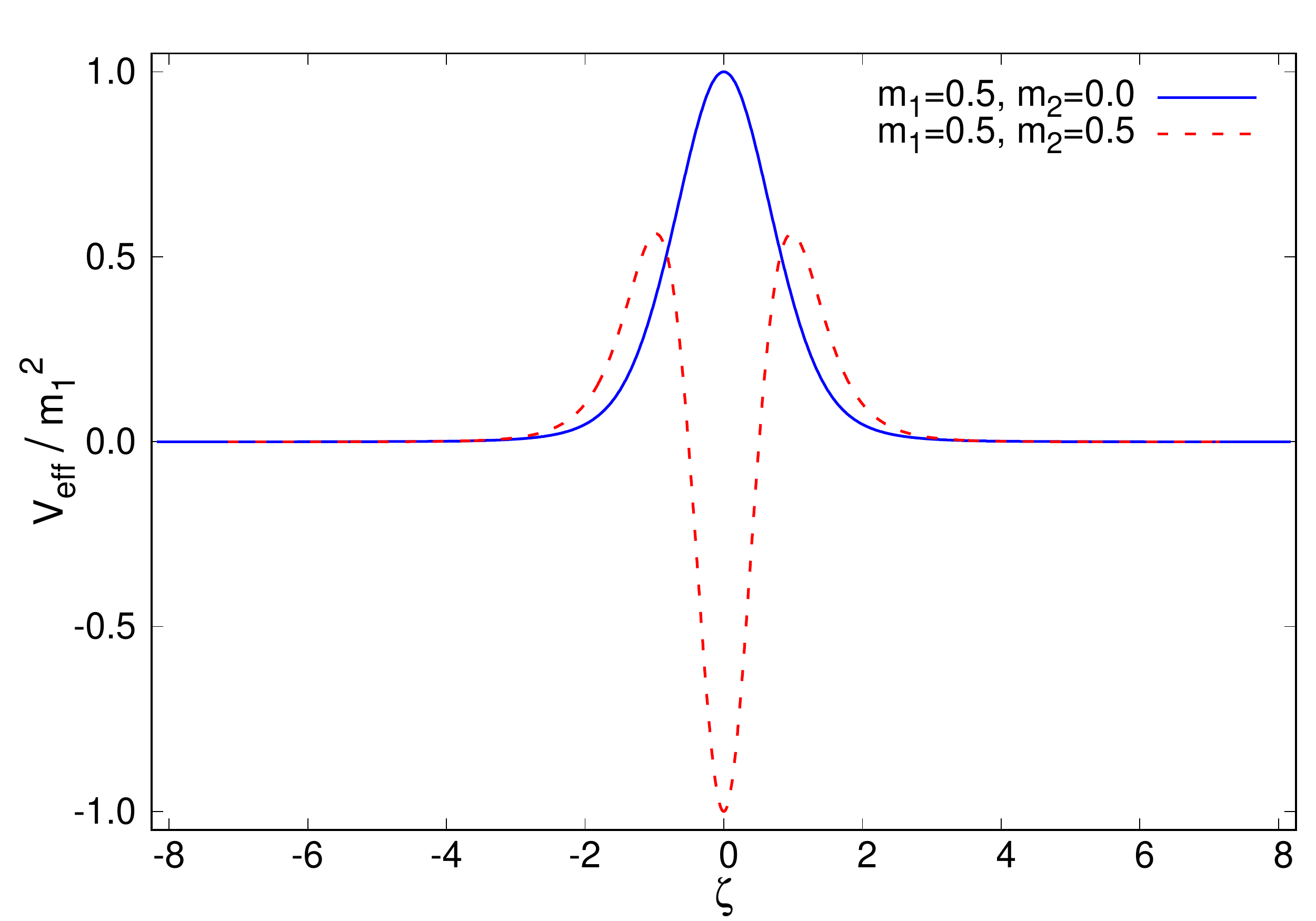}
\caption{The effective potential, obtained from the 3d simulations, plotted against $\zeta$ for the true vacuum  $Q=1$ Skyrmions at $m_1=0.5$ and $m_2=0$ and for the false vacuum $Q=1$ Skyrmions at $m_1=0.5$ and $m_2=0.5$.
}
\label{fig:pot}
\end{figure}

The potential of the Skyrme model can be adjusted to model various physical effects, for example, to construct Skyrmions with low binding energies, the pion mass term \re{pion-pot} can be supplemented with an additional term proportional to ${\rm Tr}~(\one - U)^4 $  \cite{Gillard:2015eia}. Another interesting
possibility is to consider a false vacuum
potential \cite{Dupuis:2018utr}

\be
V=-\frac{1}{4}\,\left[m_1^2\,\tr\(\one - U\)+m_2^2\,\tr\(\one - U^2\)\right] \, ,
\label{pot}
\ee
where $U=\one$ is the true vacuum. Therefore, the theory \eqref{eng_Sk} has an $SO(3)$ isospin symmetry corresponding to the field transformation $U \rightarrow O\,U\,O^{-1}$, $\forall O \in SU(2)$. If $U\to -\one$ as
${\bf x} \to \infty$, the following potential shift by $m_1^2$ is essential
\be
V \to V - m_1^2~~ \xrightarrow[{\bf x}\to \infty]~~ 0
\,.\label{shift-pot}
\ee
Below, we assume that the Skyrme field asymptotically may approach both the true and the false vacuum.  In the latter case the Skyrmions contain the true vacuum in their core. For later convenience we can introduce an effective potential (see Figure \ref{fig:pot})
\be
V_{{\rm eff}}\equiv\left\{ \begin{array}{ll} V, & U(\infty)=+\one \:\,{\rm (Skyrmions)}\\  V-m_1^2, & U(\infty)=-\one
\:\,{\rm (False \:\,vacuum \:\, Skyrmions)}\end{array} \right. \,.\label{effective}
\ee

Note that the potential \re{pot} may possess a local minimum $U_{{\rm false}}=-\one$ if, and only if $m_1^2<4\,m_2^2$. For $m_1^2\geq 4\,m_2^2$ the field configuration $U=-\one$ is a global maximum. It is not difficult to check it expanding the $SU(2)$ matrix-valued Skyrme field  $U=\one\,\phi_0+i\,\phi_a\,\tau_a$, where the quartet $(\phi_0, \phi_a)$ parametrizes the unit sphere $S^3$. Then the constraint $\phi_a^2=1-\phi_0^2$ allows us to express the potential \re{pot} as a polynomial of $\phi_0$ and find its extrema.

The variation of the Lagrangian of the Skyrme model with the potential term \re{pot} with respect to the field $U$, after some algebra yields the field equations
\be
\pa_\mu \(R^\mu +\frac{1}{4}\, \left[R_\nu,\, \left[R^\nu,\,R^\mu\right]\right]\) +
\frac{1}{8}\,m_1^2\(U-U^\dagger\)+\frac{m_2^2}{4}\,\(U^2-U^{\dagger 2}\) =0 \, .
\label{eqendb}
\ee

The asymptotic analysis of the Skyrme field becomes more simple if we make use of the expansion $U=\one\,\phi_0+i\,\phi_a\,\tau_a$, it yields
\be R_\mu  =i\,\tau_a\,\(\phi_0\,\pa_\mu \phi_a-\phi_a\,\pa_\mu \phi_0+\varepsilon_{abc}\,\pa_\mu\phi_b\,\phi_c \)
\, .\label{maurer}
\ee
Further,  we consider excitations $v_0$,\,$v_a$   of the Skyrme field around the vacua,
$U \sim (-1)^l(1-v_0)\,\one+i\,v_a\,\tau_a $. Here $\mid v_0\mid,\,\mid v_a\mid \ll 1$, and
 the value of the parameter $l=0$ corresponds to the asymptotic value of the Skyrme field $U \xrightarrow[{\bf x}\to \infty]~~\one$,  and  $l=1$ corresponds to the false vacuum asymptotic, $U \xrightarrow[{\bf x}\to \infty]~ -\one$ (if $m_1^2<4\,m_2^2$).

Note that although the $v_a$-fields  can take both positive and negative values, the $v_0$-component must be positive due to the constraint of the field to the unit sphere.  On the other hand, this constraint yields $v_a^2=1-\(1-v_0\)^2$ and since $v_0 \ll 1$, we obtain $v_a^2 \approx 2\,v_0+{\cal O}(v_0^2)$. Hence, $\pa_i v_0 \approx v_a\,\pa_i v_a +{\cal O}(v_0^2,\,v_0\,\pa_i v_0)$. Therefore,  $v_0 \ll v_a^2$ and after some algebra we arrive to the linearized form of the field equation \re{eqendb}:

\be
\pa_\mu\pa^\mu v_a + m_{{\rm eff}}^2\(l\)\,v_a=0\,, \label{asympeq}
\ee
where $m_{{\rm eff}}\(l\)$ is an effective mass parameter defined by
\be m_{{\rm eff}}\(l\) \equiv \sqrt{m_2^2+(-1)^l\,\frac{m_1^2}{4}}\,,\qquad\qquad\quad l\equiv\left\{ \begin{array}{cl} 0, & U(\infty)=+\one \\ 1, & U(\infty)=-\one
\end{array} \right. \,.\label{asympeq2}
\ee
Note that the scalar field component $\phi_0$ always remain massless while all $\phi_a$-fields have the same effective mass, which in turn can take on pure real or imaginary values. Evidently, if the Skyrme field asymptotically approached the true vacuum $(l=0)$, the equation \re{asympeq} corresponds to the usual Klein–Gordon equation and the triplet of pions has the same non-negative effective mass $m_{\rm eff}^{\rm true \:vac.}=\sqrt{m_2^2+\frac{m_1^2}{4}}$. In the massless limit ($m_1^2=m_2^2=0$) the asymptotic triplet of pion fields represent the field of three mutually orthogonal scalar dipoles, $v_a \propto r^{-2}$,  see e.g. \cite{mantonbook,Shnir:book1}.

For the false vacuum Skyrmions we have to consider three possibilities. First, for $m_{{\rm eff}}\(l=1\)=0$ the pion excitations are massless, as in the usual minimal Skyrme model without a potential.  However, the asymptotic value of the Skyrme field $U\to -\one$ now corresponds to a local maximum of the potential \re{pot}, such a massless configuration contains a bubble of the true vacuum in the interior region, the solutions may be unstable.

Secondly, as $m_{{\rm eff}}^2\(l=1\)<0$, the potential \re{pot} also possesses a global maximum  at $U=-\one$, in such a case the linearized equation \re{asympeq} corresponds to the
excitations with purely imaginary mass, i.e. the false vacuum configuration is unstable, the domain of the true vacum exponentially grows. Thirdly, as $m_{{\rm eff}}^2\(l=1\)>0$, the linearized equation \re{asympeq} describes the triplet of massive pions with the same effective mass $m_{\rm eff}^{\rm false \:vac.}=\sqrt{m_2^2-\frac{m_1^2}{4}}$.
This is probably the most interesting case, the corresponding  false vacuum Skyrmions are metastable, they can be destroyed via quantum tunneling \cite{Kobzarev:1974cp,Coleman:1977py} although the decay may be strongly suppressed.

\section{The rational map ansatz}
\label{sec:rational}
\setcounter{equation}{0}

Numerical simulations reveal that Skyrmions of higher degrees
are not spherically symmetric, they possess very geometrical
shapes \cite{Braaten:1989rg,Battye:1997qq,Houghton:1997kg}.
Rational map approximation \cite{Houghton:1997kg} gives surprisingly good approximations to the exact numerical solutions. The idea of the rational map ansatz is
to map the spheres $S^2$ centered at the origin of domain space $\mathbb{R}^3$ onto the spheres $S^2$ which
correspond to latitudes in the sphere $S^3$, the group space of the Skyrme model \cite{Houghton:1997kg}. Explicitly, let us
consider decomposition of the $SU(2)$ group element $U$ in terms of a real valued profile function
$f$ and a complex valued function $u$ \cite{mantonbook,Houghton:1997kg,luiz1, infinitesym}
\be
U = \one \,\cos f + \frac{i\,\sin f}{1+\mid u \mid^2}\,\(\begin{array}{cc} 1-\mid u \mid^2 & -2\,i\,u \\ 2\,i\,\bar{u} & -1+\mid u \mid^2 \end{array}\) \,,\label{decomposition}
\ee
The scalar components of the Skyrme field $U=\one\,\phi_0+i\,\phi_a\,\tau_a$ can be written  as
\be
\phi_0 = \cos f\,,\quad \phi_1 = \sin f\,\frac{\bar{u}+u}{1+\mid u \mid^2}\,,\quad \phi_2 = \sin f\,\frac{i\,\(\bar{u}-u\)}{1+\mid u \mid^2}\,,\quad \phi_3 = \sin f\,\frac{1-\mid u \mid^2}{1+\mid u \mid^2} \,.\label{phid}
\ee

A point on the domain space ${\mathbb R}^3 $ can be written in polar coordinates $(r,\,z,\,\bar{z})$ on the sphere $S^3$,
where $r$ is the usual radial coordinate, $z=\tan\(\theta/2\)\,e^{i\,\varphi}$ and the metric is given by
\be
ds^2=dr^2+\frac{4\,r^2}{\(1+\mid z\mid^2\)^2}\,dz\,d\bar{z} \,.
\label{holometric}
\ee
The rational map ansatz \re{decomposition} for the Skyrme field then can be written as \cite{Houghton:1997kg}
\be
u=u(z)\,\qquad\qquad \bar{u}=\bar{u}(\bar{z})\,,\qquad\qquad f=f(r) \,, \label{rational}
\ee
where $u(z)=p(z)/q(z)$ is a holomorphic rational map between the Riemann spheres  $S^2$, and $p(z)$, $q(z)$ are polynomials of $z$ with no common roots. A well-known property of the rational map is that its algebraic degree, which corresponds to the the highest degree among the polynomials $p(z)$ and $q(z)$, is equal to the topological degree of the map $u$.   It can be written in the integral representation as
\be
{\rm deg}\,u=\frac{1}{4\,\pi}\,\int_{S^2} \,d \Omega \,\vartheta  = {\rm max}\,\{{\rm deg}\,p(z),\,{\rm deg}\,q(z)\}  \label{degree} \,,
\ee
where  $\vartheta\(z,\,\bar{z}\)\equiv \(\frac{1+\mid z \mid^2}{1+ \mid u \mid^2}\)^2 \,\frac{du}{dz}\,\frac{d\bar{u}}{d\bar{z}}$ and we introduced the differential solid angle $$d\Omega=\sin \theta \,d\theta \wedge d\psi=\frac{2\,i\,dz\wedge d\bar{z}}{(1+\mid z \mid^2)^2} \, .
$$

The corresponding topological charge density ${\cal Q}$ and topological charge \eqref{chargemain} associated with the Skyrme field \re{decomposition} are respectively
\be
{\cal Q}= -\frac{f'\,\sin^2 f}{2\,\pi^2\,r^2}\,\vartheta\(z,\,\bar{z}\)\,,\qquad\qquad Q = \frac{1}{\pi}\,\left[f-\frac{1}{2}\,\sin \(2f\)\right]_{r=\infty}^{r=0}\,{\rm deg}\,u \, , \label{topcharge}
\ee
where we used \re{rational} and the definition \re{degree}. Therefore, the boundary conditions on the true vacuum configurations are $f(0)=\pi$ and $f(\infty)=0$, while for the false vacuum Skyrmions we have to impose $f(0)=2\,\pi$ and $f(\infty)=\pi$. In both cases the topological degree
of the rational map is equal to the topological charge, i.e. $Q={\rm deg}\,u$. However, the false vacuum Skyrmion contains a domain of true vacuum in its center, the configuration is classically stable \cite{Dupuis:2018utr}.

The angular part of the topological charge density \eqref{topcharge} can be writen alternatively as $\vartheta = \frac{(1+\mid z \mid)^2}{(\mid p \mid + \mid q \mid)^2}\,\mid W(z)\mid^2$, where we introduced the Wronskian \be W(z)\equiv q(z)\,p'(z)-p(z)\,q'(z) \,,\label{wronskian}\ee which is a polynomial with maximum degree $2\,\(Q-1\)$. It so follows that the topological charge density vanishes at the roots of the Wronskian, which due to $z=\tan\(\theta/2\)\,e^{i\,\varphi}$ corresponds to angular directions. Indeed, although such result is obtained in the rational map approximation, its shed light on why the isosurfaces of topological charge density of the Skyrmions possesses $2\,\(Q-1\)$ holes, at least for small values of $Q$ \cite{mantonbook}.

The rational map ansatz \re{decomposition}, \re{rational} yields the simple radial energy functional normalized by the usual factor $12\,\pi^2$
\be
E =\frac{1}{3\,\pi}\int_0^\infty dr \,\(r^2\,f^{\prime 2}+2\,Q\,\(f^{\prime 2}+1\)\,\sin^2 f+{\mathcal I} \,\frac{\sin^4 f}{r^2}+r^2\,V_{{\rm eff}}(f)\)\,,
\label{energy}
\ee
where ${\mathcal I}$ is the angular integral
$$
{\mathcal I}=\frac{1}{4\,\pi}\,\int \(\frac{1+\mid z \mid^2}{1+\mid u\mid ^2}\,\left| \frac{d u}{dz}\right|\)^4 \,\frac{2\,i\,dz\,d\bar{z}}{\(1+\mid z\mid^2\)^2}\,.
$$

For the spherically symmetric $Q=1$ Skyrmion $u=z$ and  ${\mathcal I}=1$, while for Skyrmions of higher degrees $1<B\leq 22$ the best approximation to the global minima is given by the rational map \re{decomposition} with   (see e.g. \cite{mantonbook}).
\be {\mathcal I} \approx 1.28 \,Q^2\,. \label{approx}\ee

Note that in the energy functional \re{energy}
for the false vacuum Skyrmions we used the shifted potential \re{shift-pot}, as
described above.
\be V_{{\rm eff}}=V-m_1^2\,,\qquad\qquad\quad V=m_1^2 \, \sin^2 \(\frac{f}{2}\)+m_2^2\,\sin^2 f \, .
\label{poteff}
\ee
The corresponding variational equation on the profile function $f$ is
\be
\(r^2+2\,Q\,\sin^2 f\)f''+2\,f'\,r+\sin \(2\,f\)\(Q\(f^{\prime 2}-1\)-\frac{{\mathcal I}\,\sin^2 f}{r^2}\)-\frac{r^2}{2}\frac{\delta V_{{\rm eff}}}{\delta f}=0 \,.\label{eq1}
\ee

The components of the pion triplet given in \eqref{phid} has the form $\phi_a = \sin f(r) \,\vp_a\(z,\,\bar{z}\)$, for $a=1,\,2,\,3$, where $\vp_a\(z,\,\bar{z}\)$ represents the angular dependence, which clearly decouples from the radial part. Asymptotically, we can write the perfil function as $f=l\,\pi+g$, where the $g$-field is an excitation of the vacuum $(l=0)$ or the false vacuum $(l=1)$. Therefore,  the
fluctuations of the pion fields are of the
form $v_a= (-1)^l\,g(r)\,\vp_a\(z,\,\bar{z}\)$ (see section \ref{sec:model}).
Hence, the asymptotic equation \re{asympeq} becomes
\be
r^2\,g''+2\,r\,g' - \(r^2\,m_{{\rm eff}}^2\(l\)+2\,\vartheta\(z,\,\bar{z}\)\)\,g= 0 \,,
\label{reduced}
\ee
where
we make use of the relation
$\pa_i^2 = \frac{\(1+\mid z \mid^2\)^2}{r^2}\,\pa_{z\,\bar{z}}+ \pa_r^2+\frac{2}{r}\,\pa_r$, which follows from the
explicit form of the Riemanian metric \eqref{holometric}, and $\pa_{z\,\bar{z}}\vp_a=-\frac{2\,\pa_z u\,\pa_{\bar{z}} \bar{u}}{\(1+\mid u \mid^2\)^2}\,\vp_a$. 
Note that the angular dependent term $\vp_a\(z,\,\bar{z}\)$ in the radial asymptotic equation \re{asympeq}  is always decoupled.

Multiplying the equation \re{reduced} by $\frac{1}{4\,\pi}\,d\Omega$, integrating over the $S^2$ and using \eqref{degree}, we obtain the radial asymptotic equation
\be
r^2\,g''+2\,r\,g' - \(r^2\,m_{{\rm eff}}^2\(l\)+2\,Q\)\,g= 0 \,.
\label{reduced2}
\ee
Note that for $m_{{\rm eff}}\(l\)=0$ the potential term vanishes both for the true and false vacuum Skyrmions, the radial function $g(r)$ decays asymptotically as $g\propto a/r^2$ and in both cases the asymptotic triplet of pion fields represents the field of three mutually orthogonal scalar dipoles. The equation \eqref{reduced2} can be obtained alternatively considering the asymptotic regime of \eqref{eq1} with $f=l\,\pi+g$.

Clearly, the energy of the false vacuum Skymions diverge as $m_{{\rm eff}}^2\(l\)=m_2^2 - m_1^2/4 <0$ and the effective mass parameter in the asymptotic Klein-Gordon equation \re{reduced2} becomes purely imaginary.

\section{Numeric solutions}
\label{sec:numeric1}
\setcounter{equation}{0}
To find stationary points of the energy functional
\re{eng_Sk} we implement the simulated annealing technique \cite{Hale} numerically relaxing initial field configurations, produced by the rational map approximation in a sector of topological degree $Q$.
As a consistency check, we verify that our algorithm correctly reproduces the known results for the Skyrmion configurations of the usual rescaled Skyrme model with pion mass potential \re{pion-pot},
for degrees up to $Q=6$ it agrees with previously known values of the ratio $E/Q$ within $1.0\%$ accuracy. For each solution
we evaluated the value of the topological charge $Q$,
we find that this is accurate  to within $10^{-3}$ in all simulations reported here.
Another
check of the correctness of our results was
performed by verifying that the virial relation for the Skyrme model in 3+1 dimensions between the potential, quadratic, and quartic in derivatives terms in the static energy functional, $E_2=E_4-3\,E_0$ is satisfied. Here
\be
\begin{split}
E_0  &\equiv \frac{1}{12\,\pi^2}\int d^3x \,\,V_{{\rm eff}}\,,\quad \qquad \quad
E_2\equiv \frac{1}{24\,\pi^2} \,\int d^3x\, \tr \(R_i\,R_i\)\,,\\
E_4& \equiv \frac{1}{192\,\pi^2} \,\int d^3x\,
\tr\(\left[R_i,\,R_j\right]\,\left[R_i,\,R_j\right]\)\,.
\end{split}
\ee
More precisely, following the discussion of the loosely bounded Skyrmions presented in \cite{Gillard:2015eia}, we evaluated the quantity
\be
{\cal D}=\frac{E_4-E_2-3E_0}{E_4+E_2+E_0}
\label{Derek_D}
\ee
for each solution we found. In order to have a measure of the characteristic size of Skyrmions we also introduce the root mean square (rms) radius defined by
\be \sqrt{\langle r^2 \rangle} \equiv \sqrt{\frac{1}{Q}\,\int d^3x \,r^2 \,{\cal Q}}\label{rms} \,.\ee

Note that using the virial identity we can also estimate the accuracy of the rational map approximation \eqref{rational} for which
\br
 E_2 &=& \frac{1}{3\,\pi}\int_0^\infty dr \,\(r^2\,f^{\prime 2}+2\,Q\,\sin^2 f\)\,,\nonumber\\
E_4 &=& \frac{1}{3\,\pi}\int_0^\infty dr \,\(2\,Q\,f^{\prime 2}\,\sin^2 f+{\mathcal I}\,\frac{\sin^4 f}{r^2}\)
\,,\label{energiesrational}\\
E_0 &=& \frac{1}{3\,\pi}\int_0^\infty dr \,r^2\,V_{{\rm eff}}(f) \,.\nonumber
\er
We found that the virial constraint is  satisfied with an accuracy of order of $10^{-2}$ for all solutions considered in this paper.

The results presented here used cubic grids containing $120^3$ points and Dirichlet boundary conditions  \eqref{asympeq} are
imposed, i.e. we set $U=(-1)^l\,\one$ at the edge of the lattice. The spatial grid spacing is $\Delta x = 0.08$, and consequently the spacial grid size is $R=9.52$.
Smaller grid spacing was used to study specific domains of negative values of the topological charge densities.
For comparison we also
solved one-dimensional radial equation for the $Q=1$ spherically symmetric Skyrmion on the grid with spacing $\Delta x = 0.005$ and $4001$ points, the errors in evaluation of the topological degree are less that $0.1\%$.

\subsection{False vacuum Skyrmions}
\label{sec:falsen}
\setcounter{equation}{0}
First, we considered multi-soliton solutions of the Skyrme
model with the false vacuum potential \re{pot} we discussed above. While the previous consideration of this system in
\cite{Dupuis:2018utr} was restricted to the rational map ansatz with the approximation \eqref{approx}, even for the Skyrmions of very high degrees, we
perform full 3d numerical computations to find  corresponding global minima. Evidently, these configurations do not possess spherical symmetry, the false vacuum Skyrmions have an effective mass $m_{{\rm eff}}\equiv \sqrt{m_2^2-\frac{m_1^2}{4}}$ and the binding energy of these Skyrmions is large. We found that the rational map approximation yields a very good initial approximation both for the true vacuum Skyrmions with the usual pion mass potential \re{pion-pot} and for the false vacuum Skyrmions with the potential \re{pot}.

\begin{figure}[H]
\centering
\includegraphics[scale=0.5,angle=0]{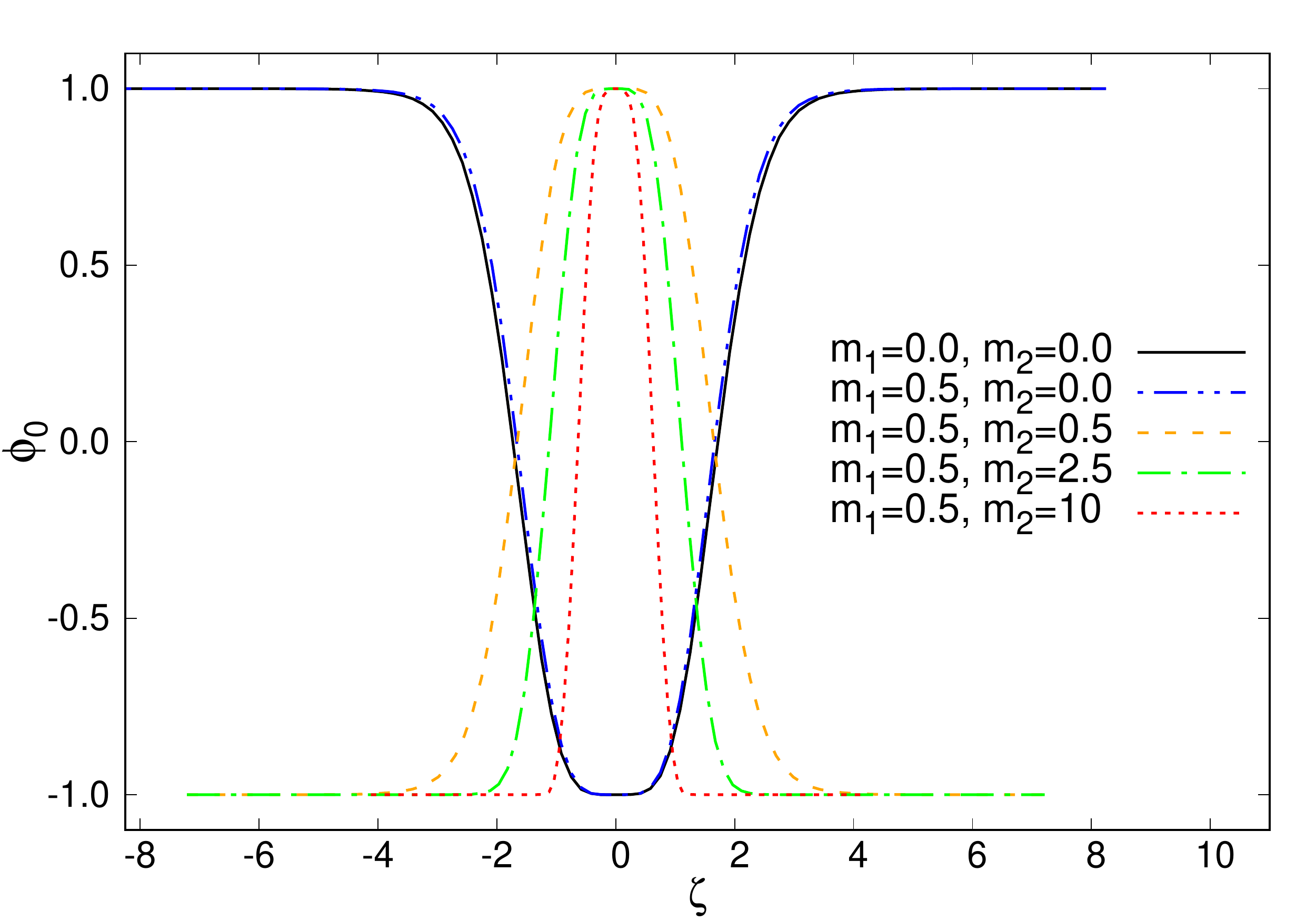}
\caption{The profile of the scalar component $\phi_0$ of the field of  $Q=4$ Skyrmion plotted against $\zeta$ for the true vacuum Skyrmions at $m_1=0,\,0.5$, $m_2=0$ and for the
false vacuum $Q=4$ Skyrmions at $m_1=0.5$ and some set of values of $m_2$.}
\label{phi0}
\end{figure}

\begin{figure}[H]
\centering
\includegraphics[scale=0.45,angle=0]{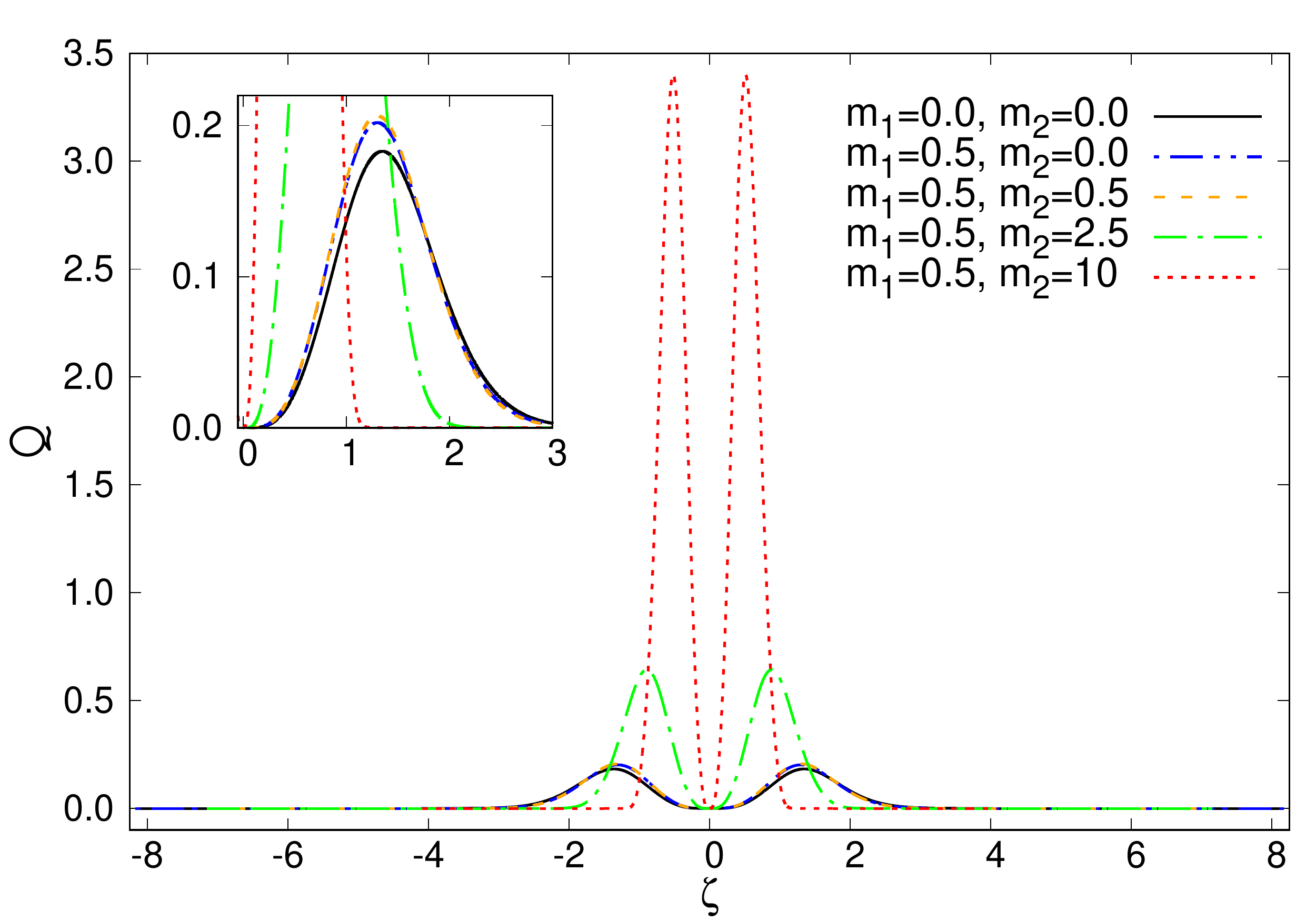}
\caption{The topological charge density $\mathcal{Q}$ plotted against $\zeta$ for the true vacuum Skyrmions at $m_1=0,\,0.5$, $m_2=0$  and for the corresponding false vacuum Skyrmions  at $m_1=0.5$ and some set of values of $m_2$.}
\label{chargedensity}
\end{figure}

\begin{figure}[H]
\centering
\includegraphics[scale=0.455,angle=0]{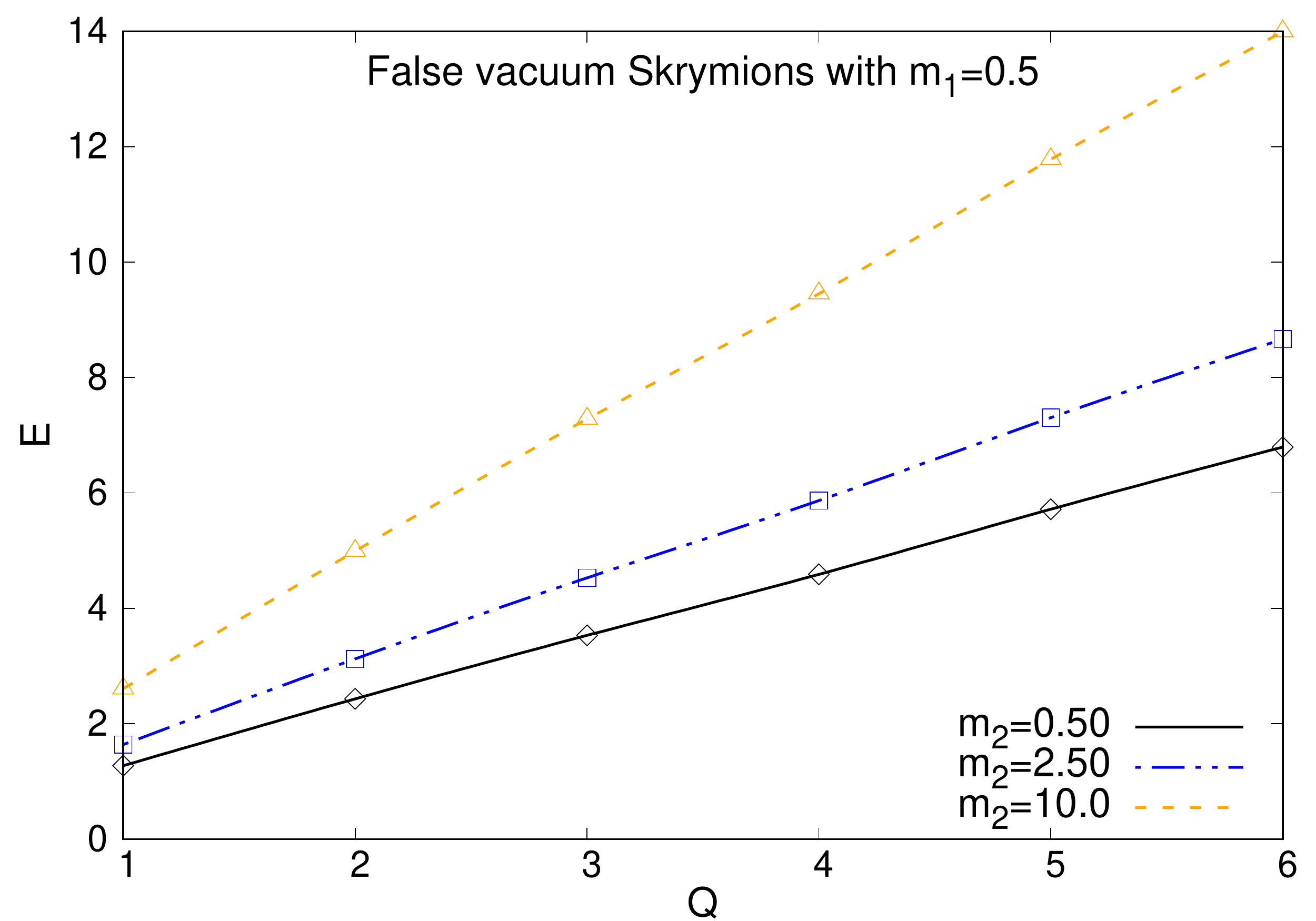}
\caption{The values of energy $E$ of the false vacuum Skyrmions, presented in the Tables \ref{tablec}-\ref{tablee},
plotted against $Q$.}       \label{figure2}
\end{figure}

\begin{figure}[H]
\centering
\includegraphics[scale=0.5,angle=0]{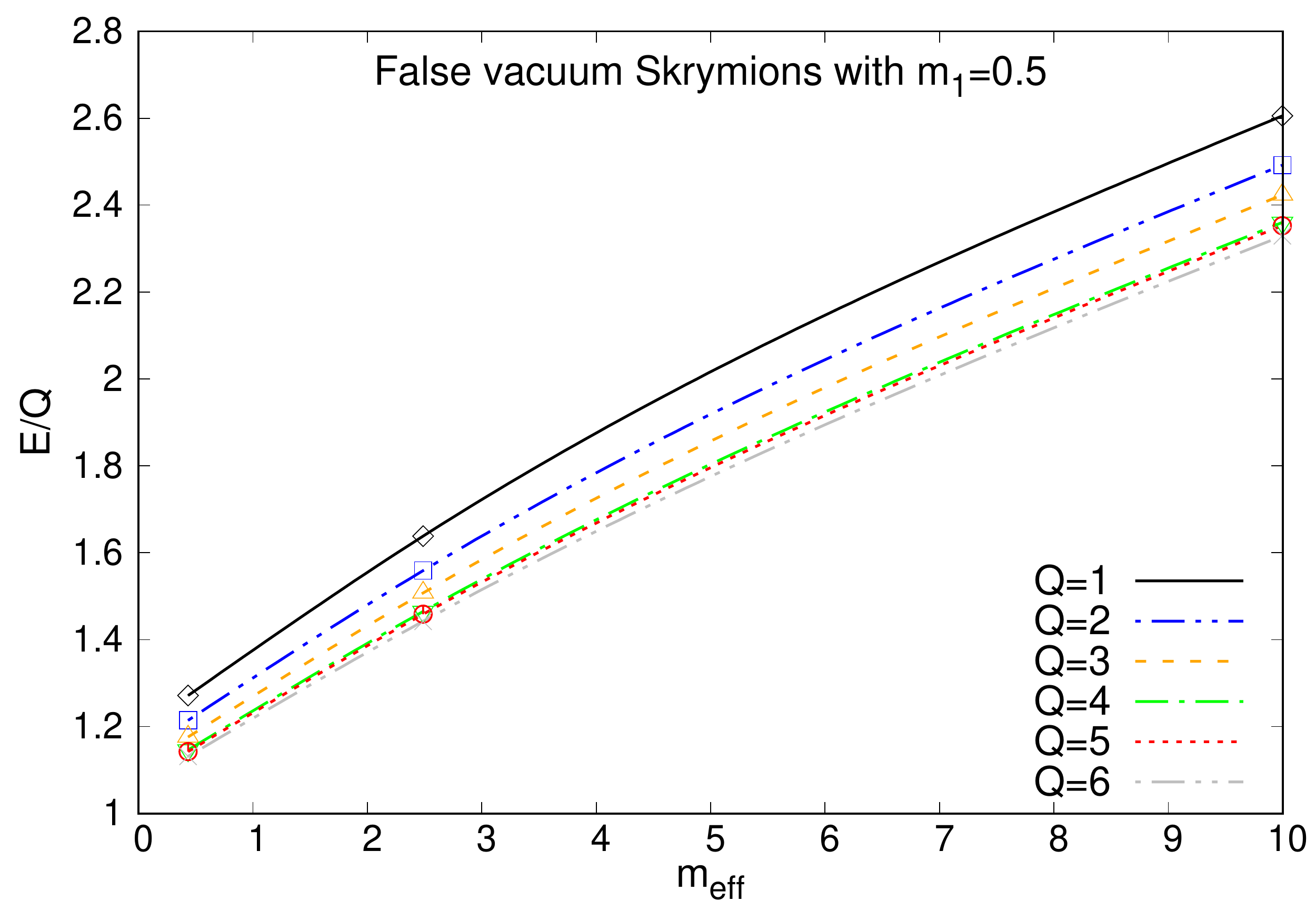}
\caption{The values of the energy per topological charge $E/Q$
of the false vacuum Skyrmions, presented in the Tables \ref{tablec}-\ref{tablee}, plotted against the effective mass $m_{{\rm eff}}\equiv \sqrt{m_2^2-\frac{m_1^2}{4}}$.}       \label{figure3}
\end{figure}

The distributions of the energy and topological charge density of the false vacuum Skyrmions depend on the effective mass $m_{{\rm eff}}$ similar to the corresponding minimizers in the model with the pion mass potential \re{pion-pot}, increase of $m_{{\rm eff}}$
yields increase of the energy density inside the core of the configuration and the size of the configuration decreases.
As an example, Figures~\ref{phi0} and \ref{chargedensity} display plots of the profiles of scalar component $\phi_0$ and topological charge density on the diagonal line through the center of the $Q=4$ configuration for some set of values of the parameters $m_1,m_2$. Such line correspond with the points where all Cartesian coordinates $(x_1,\,x_2,\,x_3)$ are equal, which can be easily parameterized by $\zeta \equiv {\rm sign}\(x_1\)\,r$. The results of our simulations are presented in Tables
\ref{tablec}-\ref{tablee} and Figures~\ref{figure2},\ref{figure3}.

\begin{table}[H]
    \caption{False vacuum Skyrmions at $m_1=0.5,\,m_2=0.5$ with topological charges $Q=1-6$: Numerically computed values of the ratio $E/Q$, Derrick constraint \eqref{Derek_D}, the rms radius \eqref{rms} and the values of the static energy $E$ and the topological degree $Q$ (num.) evaluated in full 3d numerical simulations. In addition, the first row records the data for the spherically symmetric Skyrmion $Q=1^*$ on 1d grid with $4001$ points and lattice spacing $\Delta x = 0.005$.}
        \label{tablec}
    \centering
\begin{tabular}{lccccccc}\hline
$Q$ & $E/Q$ & ${\cal D}$ & $\sqrt{\langle r^2 \rangle}$ & $E$ & $Q$ (num.)\\ \hline
$1^*$ & $1.2716$ & $-0.0016$ & $0.9655$ & $1.2716$ & $1.0000$\\
$1$ & $1.2761$ & $0.0213$ & $0.9510$ & $1.2770$ & $1.0007$ \\
$2$ & $1.2146$ & $0.0155$ & $1.3204$ & $2.4308$ & $2.0014$ \\
$3$ & $1.1761$ & $0.0186$ & $1.5510$ & $3.5301$ & $3.0015$ \\
$4$ & $1.1464$ & $0.0219$ & $1.7355$ & $4.5885$ & $4.0024$ \\
$5$ & $1.1426$ & $0.0282$ & $1.9343$ & $5.7163$ & $5.0028$ \\
$6$ & $1.1317$ & $0.0322$ & $2.0830$ & $6.7939$ & $6.0033$ \\
\hline
\end{tabular}
\end{table}

Figure \ref{figure2} demonstrates that the energies of the $Q=1-6$ false vacuum Skyrmions
 increases approximately linearly with $Q$, which is expected since the binding energy per topological charge unit $E_B\equiv E_{Q=1}-E_Q/Q$ is about $5-10\,\%$ of $E_{Q=1}$ for $Q>1$ (see Tables \ref{tablec}-\ref{tablee}). However, one can expect this pattern may change for Skyrmions of higher degrees
\cite{Battye:2004rw,Battye:2006tb}. Dependency of the ratio
$E/Q$ of these configurations on the effective mass  $m_{{\rm eff}}$ is displayed in Figure \ref{figure3}. By analogy with the corresponding curves in the model with the usual pion mass potential \re{pion-pot}, it increases with $m_{{\rm eff}}$. 

\begin{table}[H]
    \caption{False vacuum Skyrmions, the same quantities as those given above in Table \ref{tablec} for
    $m_1=0.5,\,m_2=2.5$.}
        \label{tabled}
    \centering
\begin{tabular}{lccccccc}\hline
$Q$ & $E/Q$ & ${\cal D}$ & $\sqrt{\langle r^2 \rangle}$ & $E$ & $Q$ (num.)\\ \hline
$1^*$ & $1.6384$ & $0.0001$ & $0.6288$ & $1.6384$ & $1.0000$ \\
$1$ & $1.6381$ & $-0.0025$ & $0.6332$ & $1.6406$ & $1.0015$ \\
$2$  & $1.5591$ & $-0.0025$ & $0.8764$ & $3.1228$ & $2.0029$ \\
$3$  & $1.5075$ & $-0.0020$ & $1.0259$ & $4.5277$ & $3.0034$ \\
$4$ &  $1.4650$ & $-0.0022$ & $1.1507$ & $5.8679$ & $4.0053$ \\
$5$ &  $1.4588$ & $-0.0021$ & $1.2870$ & $7.3031$ & $5.0063$ \\
$6$ &  $1.4425$ & $-0.0021$ & $1.3899$ & $8.6658$ & $6.0074$ \\
\hline
\end{tabular}
\end{table}

\begin{table}[H]
    \caption{False vacuum Skyrmions, the same quantities as those given above in Table \ref{tablec} for $m_1=0.5,\,m_2=10$.  }
        \label{tablee}
    \centering
\begin{tabular}{lccccccc}\hline
$Q$ & $E/Q$ & ${\cal D}$ & $\sqrt{\langle r^2 \rangle}$ & $E$ & $Q$ (num.)\\ \hline
$1^*$ & $2.6055$ & $0.0000$ & $0.3609$ & $2.6055$ & $1.0000$\\
$1$ & $2.6047$ & $-0.0022$ & $0.3634$ & $2.6086$ & $1.0015$ \\
$2$ & $2.4920$ & $-0.0020$ & $0.5023$ & $4.9907$ & $2.0027$ \\
$3$ & $2.4233$ & $-0.0017$ & $0.5821$ & $7.2791$ & $3.0038$ \\
$4$ & $2.3601$ & $-0.0018$ & $0.6489$ & $9.4520$ & $4.0048$ \\
$5$ & $2.3528$ & $-0.0015$ & $0.7236$ & $11.7784$ & $5.0061$ \\
$6$ & $2.3293$ & $-0.0015$ & $0.7786$ & $13.9929$ & $6.0074$ \\ \hline
\end{tabular}
\end{table}

It is instructive to visualize the solutions of the model
\re{eng_Sk} by displaying 3d isosurfaces of the topological charge densities ${\cal Q}$, see e.g. \cite{mantonbook}. However, variation of the mass parameter significantly affects the characteristic size of the field configuration. In our simulations, for each solution we found numerically on a discretized 3d grid, we introduce an $\alpha$-function, which is defined as a sum of the edges of the elementary parallelepiped in the domain of the lattice that contains an isosurface of the topological density. Explicitly, $\alpha$ is a function of $m_1$, $m_2$, $l$, ${\cal Q}$ and $Q$, i.e. $\alpha=\alpha\(Q,\,m_1,\,m_2,\,l,\,{\cal Q}\)$.
Using this function, we can scale an isosurface of the topological density plotted for some value of the density ${\cal Q}$, to make a visual correspondence with a different
plot for another value of ${\cal Q}$. In other words, we can define a ``zoom factor''  between the corresponding isosurfaces
of the topological charge densities as, for example,
\be
\kappa(Q,\,m_2,\,{\cal Q}) \equiv 1-\alpha(Q,\,0.5,\,m_2,\,1,\,{\cal Q})/\alpha(Q,\,0.5,\,0,\,0,\,0.08)\, ,
\label{zoom}
\ee
where, as a reference point, we used the function $\alpha(Q,\,0.5,\,0,\,0,\,0.08)$, it corresponds to the usual massive Skyrmions with $m_1=0.5$.

\begin{figure}[H]
\centering
\includegraphics[scale=0.18,angle=0]{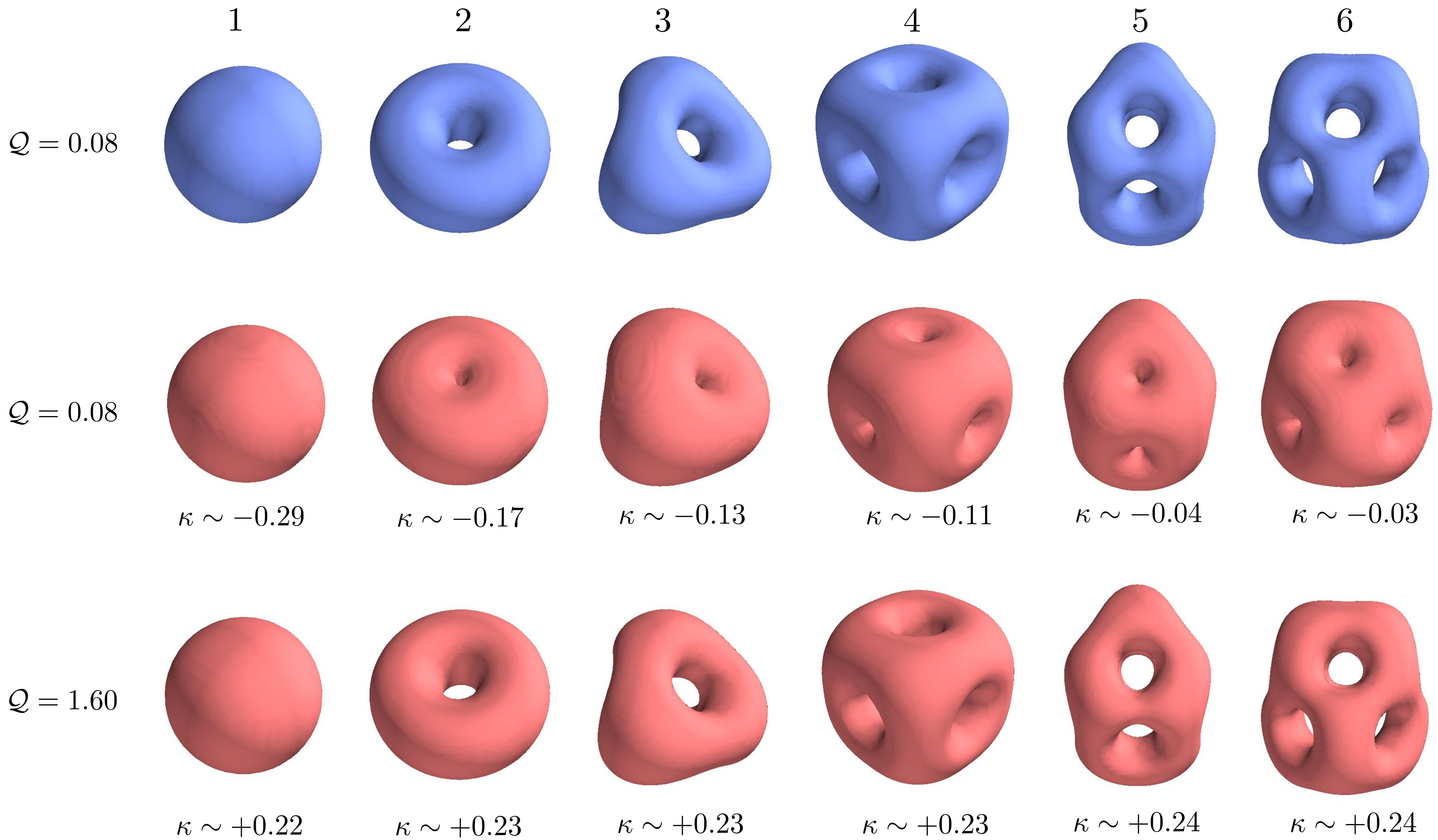}
\caption{Three sets of charge density isosurfaces of the Skyrmions degrees  $Q=1$ to $Q=6$, from left to right. The upper row displays the true vacuum massive Skyrmions (blue figures) for $m_1=0.5$ and $m_2=0$ at  ${\cal Q}=0.08$. The second and the third rows displays false vacuum Skyrmions (red figures) for $m_1=0.5$ and $m_2=10$ at ${\cal Q}=0.08$ and ${\cal Q}=1.6$, respectively,
rescaled to be approximately the same size. The zoom factor $\kappa$ is defined by \re{zoom}.}
\label{iso}
\end{figure}

Figure \ref{iso} displays the corresponding rescaled isosurfaces of the charge densities for the usual true vacuum massive Skyrmions (blue figures, upper row) for $m_1=0.5$ and $m_2=0$, and the false vacuum Skyrmions (red figures, middle and bottom rows) for $m_1=0.5$ and $m_2=10$, respectively. Clearly, geometrical shapes of the global minimizers in the Skyrme model with the true vacuum potential \re{pion-pot} and with the false vacuum potential \re{pot} are very similar.

\section{Regions of negative topological charge density}
\label{sec:negative}
\setcounter{equation}{0}

A peculiar feature of the Skyrmions is that they allow for existence of regions of negative topological charge density
\cite{Houghton:2001fe,Foster:2013bw}. It was argued that these regions are associated with singularities of the non-holomorphic rational map, this conjecture was supported by direct numerical
simulations for the tetrahedrally symmetric minimum-energy $Q=3$ Skyrmion \cite{Foster:2013bw}. It was shown that the regions of negative topological charge density of this configurations represent a tiny dual tetrahedron at the center of the configuration, with four small tubes smoothly joining it up and
further, passing though the faces of the tetrahedron of positive topological density. It was pointed out that in the model with the usual pion mass term \re{pion-pot}, the tubes become more pronounced \cite{Foster:2013bw}. It was also verified that $Q=2$ and $Q=4$ Skyrmion do not support any regions of negative baryon density \cite{Houghton:2001fe,Foster:2013bw,Leese:1993mc}.

Our aim now is to extend this study to the Skyrmions of higher degrees $Q=5,6$ also considering the model with false vacuum potential \re{pot}. Since the regions of the negative topological density are  very tiny, we have to refine our numerical algorithm to truly capture these domains.  First, we construct a Skyrmion solution on a  a lattice with $120^3$ points, with spacing $\Delta x \sim 0.08$, and then we select some particular rectangular region of the lattice. Second, we use polynomial three-dimensional interpolation to multiply the number of points inside and at the border of this small sub-lattice, by typically factor of four, reducing the lattice spacing, correspondingly. The polynomial expansion of the fields on the new sub-lattice is used to generate new input data.  Finally, we repeat the simulated annealing algorithm inside the new lattice using the fixed boundary conditions. We can repeat this procedure multiple times to investigate very tiny structures, like the regions of the negative topological densities inside the core of the Skyrmions.

\begin{figure}[H]
\centering
\includegraphics[scale=0.24,angle=0]{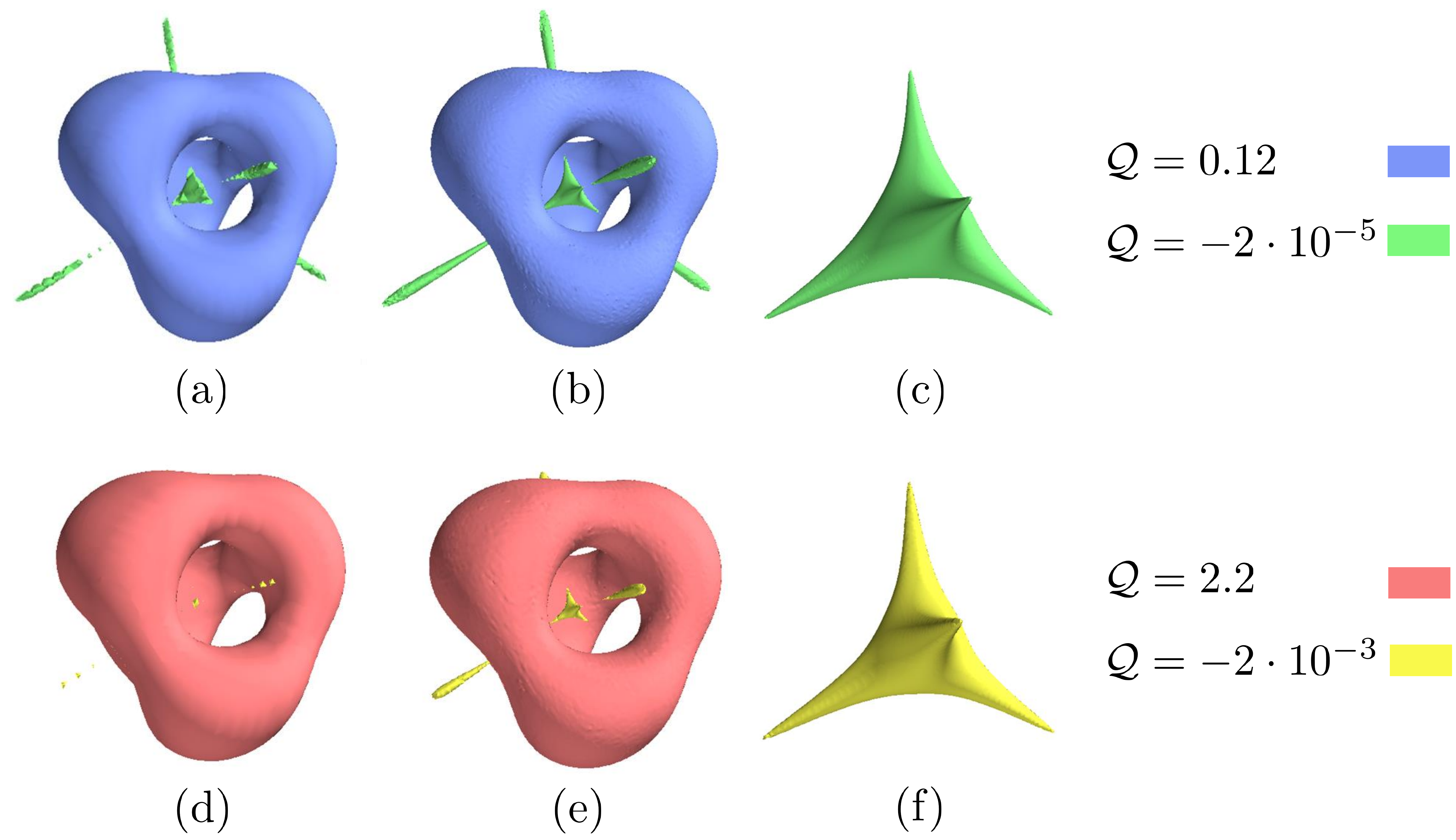}
\caption{Isosurfaces of the topological charge density for $Q=3$ Skyrmions.  Upper row:  usual massive Skyrme model potential at $m_1=0.5$ and $m_2=0$. Bottom row: the model with the false  vacuum potential \re{pot} at $m_1=0.5$ and $m_2=10$. The lattice spacing in  the simulations for the plots (a)-(c) are  $\Delta x= 0.08,\,0.02,\,0.005$, and for the plots (d)-(e) are  $\Delta x= 0.04,\,0.01,\,0.0025$, respectively.}      \label{Q=3iso}
\end{figure}

First, we revisit the $Q=3$ Skyrmion. Considering the usual  model with the pion mass potential \re{pion-pot}, we find numerically the dual tetrahedron of negative topological density about the origin,
the corners of the tetrahedron  are linked to the four tubes  \cite{Foster:2013bw}. Consequent decrease of the lattice spacing allows to refine the shapes of these regions, the values of the charge density presented in Table \ref{tablenegative} are in reasonable agreement with results reported in \cite{Foster:2013bw}. Figure \ref{Q=3iso} displays the isosurfaces of the topological charge density of the $Q=3$ Skyrmions. Clearly, increase of the resolution reveals very fine structure of the domains of negative charge density.

For the usual Skyrme model the value of the field is close to the anti-vacuum, $U\to -\one$ at the center of the configuration,
and it is taking the vacuum value on the boundary. For the model with the false vacuum potential \re{pot} the situation is inverse, see Figure \ref{phi0}. However, the structure of the domains of the negative charge density for the tehrahedral $Q=3$ Skyrmions remains the same, see  Figure \ref{Q=3iso}.  Our numerical results indicate that the presence of the these regions do not destabilize the false vacuum Skyrmions, for all range of values of the effective mass.

\begin{table}[H]
    \caption{The effective mass $m_{{\rm eff}}=\sqrt{m_2^2+l \,\frac{m_1^2}{4}}$, the minimal (${\cal Q}_-$) and the maximal (${\cal Q}_+$) values of the topological density for the tetrachedral $Q=3$ Skyrmions $(l=1)$  and false vacuum Skyrmions $(l=-1)$.}
        \label{tablenegative}
    \centering
\begin{tabular}{cccccc}\hline
Type & $m_1$ & $m_2$ & $m_{{\rm eff}}$ & ${\cal Q}_-$ $\(10^{-3}\)$ & ${\cal Q}_+$\\ \hline
Standard $Q=3$ Skyrmion & $0.0$ & $0.0$ & $0.0000$ & $-3.50$ & $0.245$ \\
Massive $Q=3$ Skyrmion & $0.5$ & $0.0$ & $0.2500$ & $-3.65$ & $0.255$ \\
False $Q=3$ Skyrmion & $0.5$ & $0.5$ & $0.4330$ & $-3.56$ & $0.262$ \\
False $Q=3$ Skyrmion & $0.5$ & $2.5$ & $2.4875$ & $-6.99$ & $0.790$ \\
False $Q=3$ Skyrmion & $0.5$ & $10.0$ & $9.9969$ & $-15.22$ & $4.036$ \\
\hline
\end{tabular}
\end{table}

It was pointed out, that the occurrence of the regions of negative topological charge density is related with zeros of the Wronskian \eqref{wronskian} associated to the rational map \cite{Houghton:2001fe,Foster:2013bw}. More precisely,
the zeros of the distribution of the topological charge density correspond to the folding of the Skyrme field, which is a map between the spheres $S^3$. It was observed, however, that
there is no regions of negative charge density for the axially symmetric $Q=2$ Skyrmion and for the $Q=4$ with cubic symmetry \cite{Houghton:2001fe,Foster:2013bw,Leese:1993mc}. Our numerical simulations confirm this result.
It was also pointed out \cite{Houghton:2001fe} that the folding structure of the holomorphic rational map of the $Q=5$ Skyrmions may give rise to the regions of negative charge density associated with  zeros of the Jacobian matrix of the Skyrme map,
but this was never observed in numerical calculations.

\begin{figure}[H]
\centering
\includegraphics[scale=0.22,angle=0]{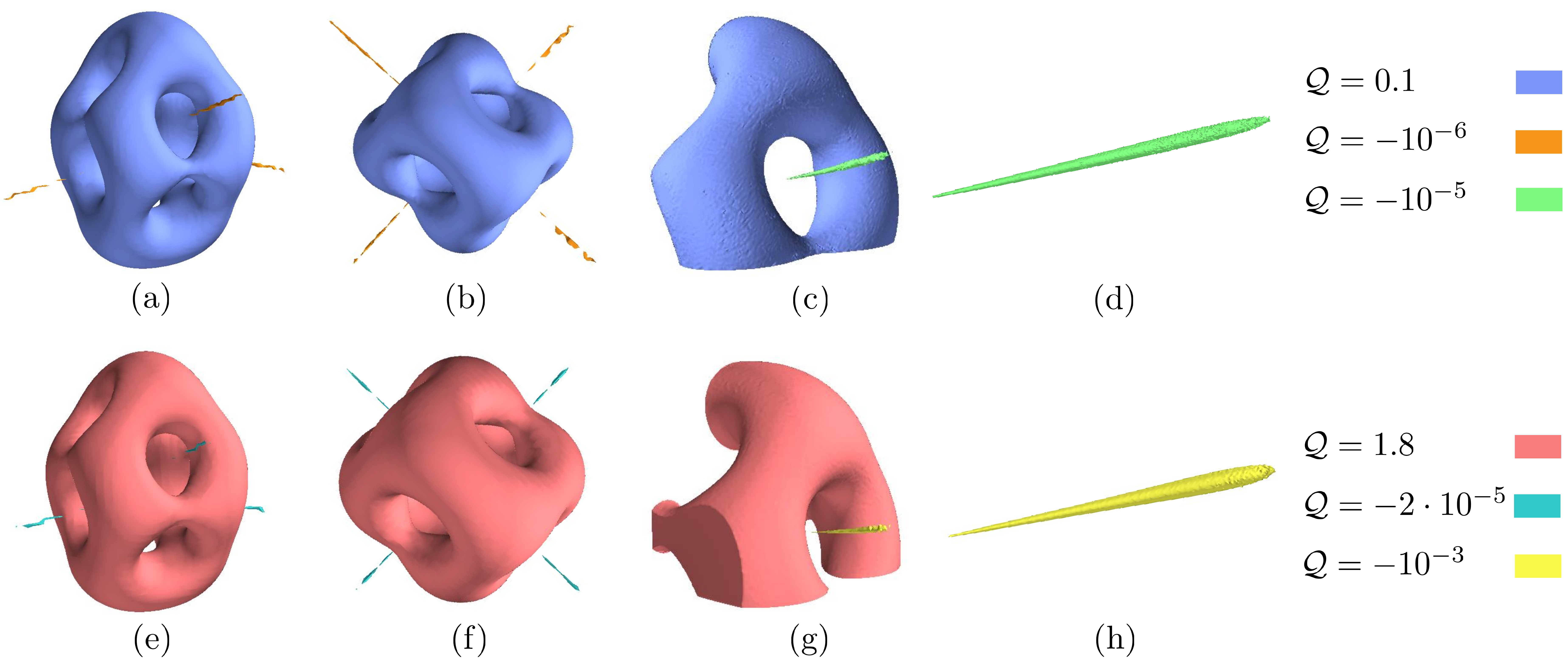}
\caption{Isosurfaces of the topological charge density for $Q=5$ Skyrmions. Upper row:  usual massive Skyrme model with the potential \re{pot} at $m_1=0.5$ and $m_2=0$. Bottom row: the model with the false  vacuum potential \re{pot} at $m_1=0.5$ and $m_2=10$. The lattice spacing in  the simulations for the plots (a)-(d)  are  $\Delta x= 0.08,\,0.08,\,0.02,\,0.005$, and for the plots (e)-(h) are  $\Delta x=0.04,\, 0.04,\,0.01,\,0.0025$, respectively. Clearly, we plot only one isosurface of negative topological charge density in each plot.}
    \label{Q=5iso}
\end{figure}

In order to check this hypothesis numerically, we perform a  detailed study of the global minimizers in the sectors of topological degrees $Q=5$ and $Q=6$. A best approximation to the
$D_{2d}$-symmetric $Q=5$ Skyrmion is given by the holomorphic rational map \re{rational}
\be
u(z)=\frac{z\,(z^4 + b\, z^2 +a)}{a \,z^4 - b \,z^2 +1}
\,,\label{map5}\ee
where $a,b$ are two real parameters. The rational map is minimized when $a=3.07, b=3.94$,  it yields a polyhedron constructed from
four pentagons and four quadrilaterals \cite{Manton:2000kj}. Since the Wronskian \eqref{wronskian} associated with \eqref{map5} possesses eight roots, it follows that this map gives rise to eight singular rays, which start at the origin. Our numerical anylyse of the minimum energy $Q=5$ Skyrmion in the model with pion mass potential \re{pion-pot} reveals four tiny regions of negative topological density, see Figure~\ref{Q=5iso}.

\begin{figure}[H]
\centering
\includegraphics[scale=0.23,angle=0]{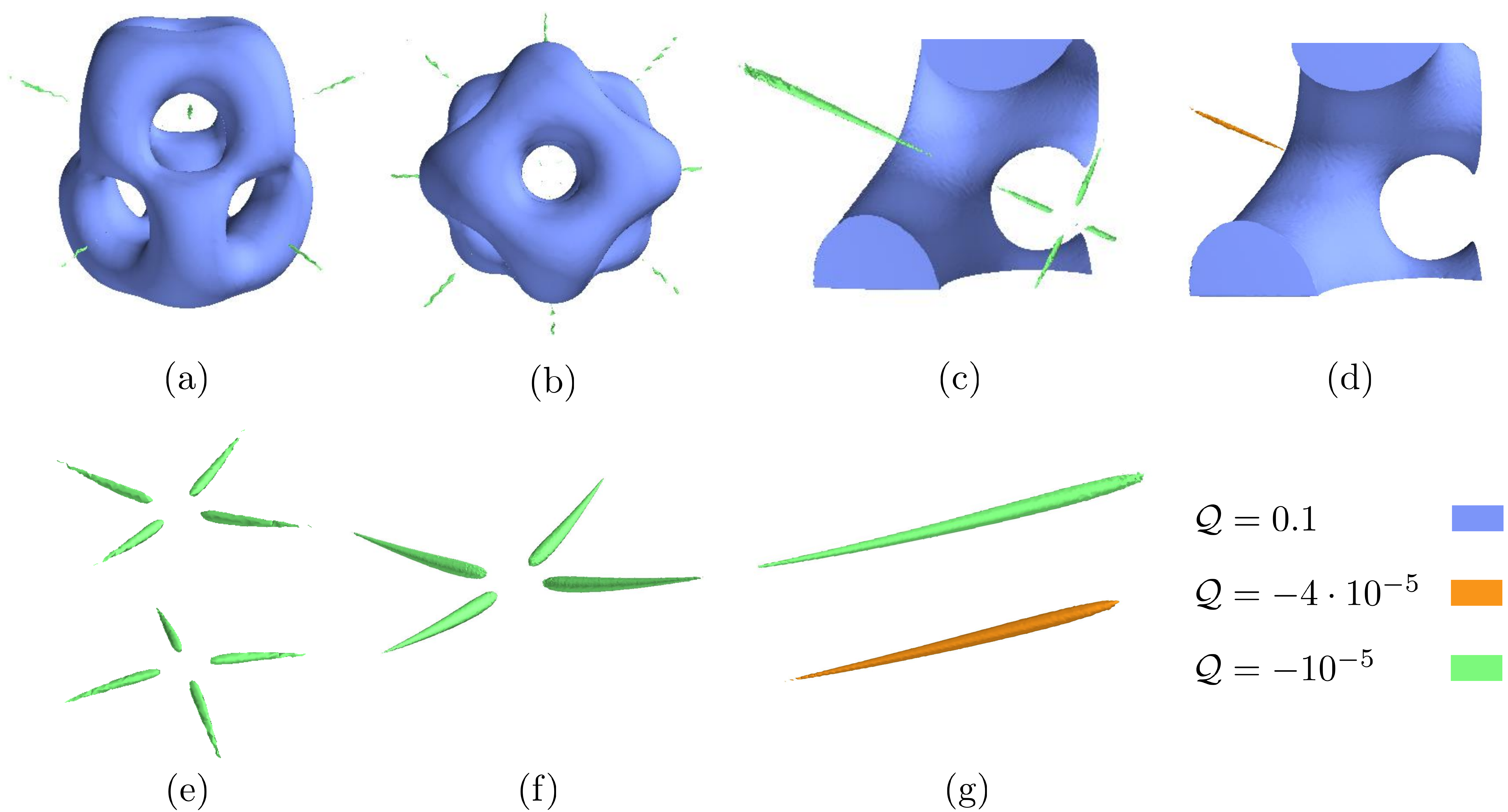}
\caption{Isosurfaces of the topological charge density for $Q=6$ Skyrmions in the usual massive Skyrme model with the potential \re{pot} at $m_1=0.5$ and $m_2=0$. The lattice spacing in  the simulations for the plots from  (a) to (g) are  $\Delta x= 0.08,\,0.08, 0.02,\,0.02,\,0.02,\,0.005,\,0.005$, respectively.}      \label{Q=6iso}
\end{figure}

The polyhedral $Q=6$ Skyrmion can be constructed via the $D_{4d}$ symmetric rational map \cite{Manton:2000kj}
$$
u(z)=\frac{z^4 +a}{z^2\,(a\,z^4 +1)}\, ,
$$
where the parameter of the map has to be taken as $a=0.16 \,i$ to minimize the energy. This map gives rise to ten singular rays, which start at the origin. The $Q=6$ configuration can be regarded as a bounded system of two Skyrmions of charges $Q= 4$ and a $Q= 2$, see Figure~\ref{Q=6iso}. Interestingly, we found 8 small tubes of negative charge density,
both in the usual Skyrme model and in the model with false vacuum potential \re{pot}.

\begin{figure}[H]
\centering
\includegraphics[scale=0.23,angle=0]{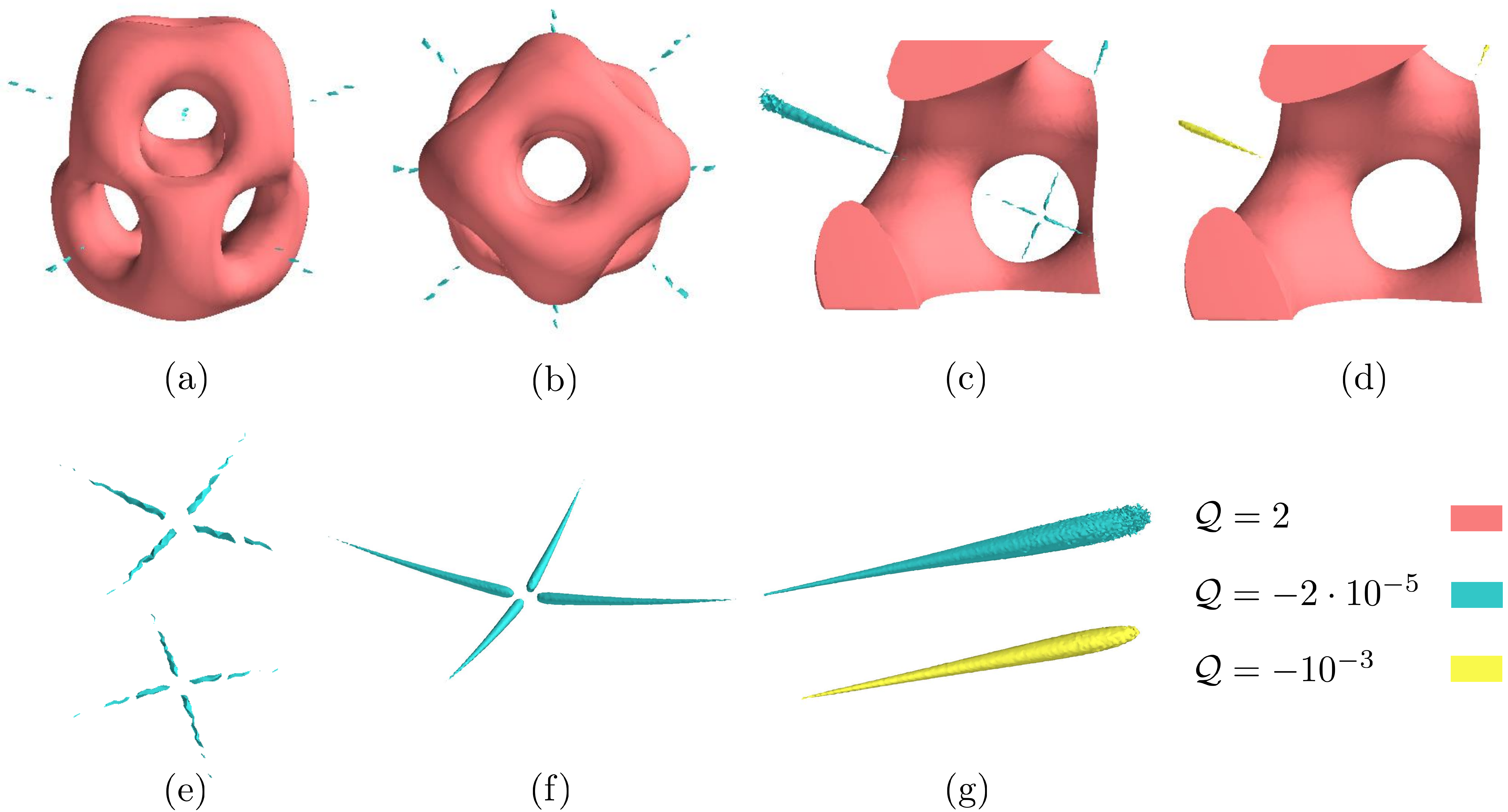}
\caption{The isosurfaces of topological charge density of the  $Q=6$ Skyrmions in the model with the false  vacuum potential \re{pot} at $m_1=0.5$ and $m_2=10$. The lattice spacing in the simulations for the figures from (a) to (g) are $\Delta x= 0.04,\,0.04, 0.01,\,0.01,\,0.01,\,0.0025,\,0.005$,  respectively. 
.}
\label{Q=6iso2}
\end{figure}

\section*{Conclusion}

Our motivation for this study is twofold. On the one hand, we want to extend the analysis of the paper \cite{Dupuis:2018utr}, to the global minimizers of the Skyrme model with the false vacuum potential \re{pot}. On the other hand, we studied the structure of the small domains of the negative topological charge densities for
Skyrmions of degrees higher than three.
We have performed fully three-dimensional numerical relaxations of Skyrmions with topological charges from $Q=1$ to $Q=6$ both
in the Skyrme model with the conventional
pion mass term included and in the model with generalized  potential \re{pot} which admits false vacuum Skyrmions. The rational map parameterization was used to generate initial data in both models. Our calculations show that, as the effective mass $m_{{\rm eff}}$ remains positive, the shapes of the soliton solutions in both models are  qualitative similar, the effective mass is playing the role of the
pion mass parameter. The false vacuum Skyrmions are metastable,  they contain a domain of true vacuum inside the core. These configurations are classically stable, however they can decay via quantum tunnelling.
On the other hand, presence of the domains of true vacuum may induce instability of the colliding false vacuum Skyrmions, as compared with scattering of the usual Skyrmions \cite{Foster:2015cpa}. Similar effect may be observed for classically isorotating Skyrmions \cite{Battye:2014qva} and baby Skyrmions \cite{Battye:2013tka,Halavanau:2013vsa} in the model with a false vacuum potential.

We also explored numerically very small regions of negative topological density which appear for the Skyrmions of degress $Q=3,5,6$ both in the model with pion mass potential and for the false vacuum Skyrmions. We confirm that these regions are associated with singularities in the rational map ansatz. Our numerical full field simulations verified previous conclusions
\cite{Houghton:2001fe,Foster:2013bw,Leese:1993mc} that the regions of negative charge density do not appear for the Skyrmions of degrees $Q=2$ and $Q=4$.

Obviously it is important to study the quantum decay rate of the metastable false vacuum Skyrmions. This task  may be performed beyond the thin wall approximation by applying advanced numerical methods, see e.g. \cite{Shkerin:2021rhy}. We believe it will be interesting to study if the regions of negative charge density may catalyse the vacuum decay. Extending this analyse to the Skyrmions of higher degrees also can be an interesting problem. Another direction for future study is to consider extended Skyrme model with sixtic term \cite{Adam:2010fg,Neto:1994bu,Floratos:2001ih} and a false vacuum potential.

Ya.S. gratefully acknowledges the support by the Ministry of
Education of Russian Federation, project No FEWF-2020-003.  The
computations in this paper were run on the "GOVORUN" cluster
supported by the LIT, JINR.



\end{document}